\documentclass[usenatbib]{mnras}

\usepackage{here}
\usepackage{multicol}
\usepackage{amssymb}
\usepackage{color}
\usepackage{lscape}
\usepackage[T1]{fontenc}
\usepackage{ae,aecompl}

\usepackage{graphicx} 
\usepackage{amsmath}	
\usepackage{amssymb}	

\title[Herschel protocluster survey]{Herschel protocluster survey: A search for dusty star-forming galaxies in protoclusters at $z=2-3$}

\author[Y. Kato et al.]{
\parbox[t]{\textwidth}{\vspace{-1cm}
Y. Kato,$^{\! 1,2}$\thanks{E-mail:kato.yu@nao.ac.jp} Y. Matsuda,$^{\! 1,3}$ Ian Smail,$^{\! 4,5}$ A. M. Swinbank,$^{\! 4,5}$ B. Hatsukade,$^{\! 1}$ H. Umehata,$^{\! 6, 7}$ I. Tanaka,$^{\! 8}$ T. Saito,$^{\! 1}$ D. Iono,$^{\! 1,3}$ Y. Tamura,$^{\! 7}$  K. Kohno,$^{\! 7,9}$ D. K. Erb,$^{\! 10}$ B. D. Lehmer,$^{\! 11}$ J. E. Geach,$^{\! 12}$ C. C. Steidel,$^{\! 13}$ D. M. Alexander,$^{\! 4}$ T. Yamada$^{\! 14}$ and T. Hayashino$^{\! 15}$}
\\\\
$^{1}$ National Astronomical Observatory of Japan, 2-21-1 Osawa, Mitaka, Tokyo, 181-8588, Japan\\
$^{2}$ Department of Astronomy, Graduate school of Science, The University of Tokyo, 7-3-1 Hongo, Bunkyo-ku, Tokyo 133-0033, Japan\\
$^{3}$ Department of Astronomy, School of Science, The Graduate University for Advanced Studies (SOKENDAI), Osawa, Mitaka, Tokyo\\
181-8588, Japan\\
$^{4}$ Centre for Extragalactic Astronomy, Department of Physics, Durham University, South Road, Durham, DH1 3LE, UK\\
$^{5}$ Institute for Computational Cosmology, Durham University, South Road, Durham DH1 3LE, UK\\
$^{6}$ European Southern Observatory, Karl-Schwarzschild-Str. 2, D-85748 Garching, Germany\\
$^{7}$ Institute of Astronomy, School of Science, The University of Tokyo, 2-21-1 Osawa, Mitaka, Tokyo 181-0015, Japan\\
$^{8}$ Subaru Telescope, National Astronomical Observatory of Japan, 650 North A’ohoku Place, Hilo, HI 96720, USA\\
$^{9}$ Research Center for the Early Universe, The University of Tokyo, 7-3-1 Hongo, Bunkyo, Tokyo 113-0033\\
$^{10}$ Center for Gravitation, Cosmology and Astrophysics, Department of Physics, University of Wisconsin Milwaukee 3135\\ North Maryland Avenue, Milwaukee, Wisconsin 53211, USA\\
$^{11}$ Department of Physics, University of Arkansas, 226 Physics Building, 835 West Dickson Street, Fayetteville, AR 72701, USA\\
$^{12}$ Centre for Astrophysics Research, Science \& Technology Research Institute, University of Hertfordshire, Hatfield, AL10 9AB, UK\\
$^{13}$ California Institute of Technology, MS 249-17, Pasadena, CA 91125, USA\\
$^{14}$ Institute of Space Astronautical Science, Japan Aerospace Exploration Agency, Sagamihara, Kanagawa 252-5210, Japan\\
$^{15}$ Research Center for Neutrino Science, Tohoku University, Sendai, Miyagi, 980-8578, Japan}

\date{Accepted 2016 May 19. Received 2016 May 16; in original form 2016 March 18}

\pubyear{2016}

\begin{document}
\label{firstpage}
\pagerange{\pageref{firstpage}--\pageref{lastpage}}
\maketitle

\begin{abstract}
We present a \textit{Herschel}/SPIRE survey of three protoclusters at $z=2-3$ (2QZCluster, HS1700, SSA22).  Based on the SPIRE colours ($S_{350}/S_{250}$ and $S_{500}/S_{350}$) of 250~$\mu$m sources, we selected high redshift dusty star-forming galaxies potentially associated with the protoclusters. In the 2QZCluster field, we found a 4$\sigma$ overdensity of six SPIRE sources around 4.5$'$ ($\sim$ 2.2 Mpc) from a density peak of H$\alpha$ emitters at $z=2.2$.  In the HS1700 field, we found a 5$\sigma$ overdensity of eight SPIRE sources around 2.1$'$ ($\sim$ 1.0 Mpc) from a density peak of LBGs at $z=2.3$. We did not find any significant overdensities in SSA22 field, but we found three 500~$\mu$m sources are concentrated $3'$ ($\sim$1.4 Mpc) east to the LAEs overdensity. If all the SPIRE sources in these three overdensities are associated with protoclusters, the inferred star-formation rate densities are 10$^3-10^4$ times higher than the average value at the same redshifts. This suggests that dusty star-formation activity could be very strongly enhanced in $z\sim2-3$ protoclusters. Further observations are needed to confirm the redshifts of the SPIRE sources and to investigate what processes enhance the dusty star-formation activity in $z\sim2-3$ protoclusters. 
\end{abstract}




\section{Introduction}
The central regions of local clusters are dominated by passive early-type ellipticals and spheroidals, their stellar populations are old, with inferred formation redshifts of $z\gtrsim2$ \citep[e.g.,][]{1997ApJ...483..582E}. High-redshift dusty star-forming galaxies (DSFGs) are  strongly star-forming galaxies (SFR $\gtrsim 100-1000$~$\rm{M_{\odot}}$\ yr$^{-1}$) and have been proposed to the precursors of present-day ellipticals in local clusters \citep[e.g.,][]{1998ApJ...500...75L,1998ApJ...507L..21S, 2001A&A...378...70L, 2013ApJ...772..137I}. Large hydro-dynamical simulations and galaxy formation models predict intense star formation could be detectable as concentrations of DSFGs in $z\gtrsim2$ protoclusters \citep[e.g.,][]{2015MNRAS.450.1320G}. Indeed, there have been reports of a reversal of the SFR-density relation \citep[e.g.,][]{2007A&A...468...33E,2010ApJ...719L.126T}, which is increasing SFR with increasing local density at $z\gtrsim1$. Protoclusters at $z\gtrsim2$ are thus unique laboratories to explore bursting star-formation in a critical epoch of galaxy formation \citep{2016arXiv160304437C}.

A numbers of studies have confirmed the presence of DSFGs in $z<2$ galaxy clusters \citep[e.g.,][]{2013ApJ...779..138B, 2014ApJ...782...19S, 2015ApJ...806..257M, 2015ApJ...809..173W}, while studies confirming DSFGs at $z\gtrsim2$ protoclusters are also progressing \citep[e.g.,][]{2009Natur.459...61T,2014MNRAS.439.1193C,2014MNRAS.440.3462U,2015ApJ...815L...8U,2015ApJ...808L..33C}. Some surveys to search for DSFGs at far-infrared wavelengths have focused on radio galaxy fields \citep[e.g.,][]{2003Natur.425..264S,2013MNRAS.436.2505V, 2014MNRAS.437.1882R, 2014A&A...570A..55D}. Radio galaxies are thought to be tracers of large scale structures, and some fraction of $z\gtrsim2$ protoclusters around radio galaxies indeed appear to have experienced bursting dusty star-formation related to DSFGs \citep[e.g.,][]{2000ApJ...542...27I}. However, such regions may be biased by the end to host a currently accreting Super Massive Black Hole (SMBH) and so we also need to explore DSFGs large-scale structures selected by other techniques at $z\gtrsim2$ to investigate bursting star-formation more generally. \citet{2014MNRAS.439.1193C} investigated $\sim$ 90 deg$^2$ sky observed as part of the HerMES survey with \textit{Planck} and \textit{Herschel} to search for clusters undergoing dusty star-formation. They found four candidate clusters, and for all four cases they found evidence of galaxy clusters with red-sequence based on optical/NIR data. The star-formation rate density of these at $z\gtrsim2$ are four order of magnitudes higher than the cosmic averaged values. But is this also true for dusty star-formation in known optical/UV-selected galaxy clusters at $z\gtrsim2$?

In this paper, we report a result of observations with the Spectral and Photometric Imaging Receiver \citep[SPIRE;][]{2010A&A...518L...3G} on the  \textit{Herschel\ Space\ Observatory} \citep[\textit{HSO}; ][]{2010A&A...518L...1P} for three protoclusters at $z=2-3$ (2QZCluster, HS1700, and SSA22 at $z=2.2$, 2.3, and 3.1, respectively). The three protoclusters have filamentary, large scale structures of rest-frame UV to optical selected galaxies. The structure of this paper is following.  In \S 2 we introduce the  \textit{Herschel}/SPIRE observations, the data processing, and our targeted fields. In \S 3 we present source detection methods, number counts, and SPIRE colour selection. In \S 4 and \S5 we present our results, discussion, and then summarize our main findings. We use the following cosmological parameters: $\Omega_m = 0.3,\ \Omega_\Lambda = 0.7,\ h = 0.7$. In this cosmology, the Universe is 2.9, 2.8, and  2.0~Gyr old and 1.0$''$ corresponds to 8.3, 8.2 and 7.6~kpc in physical length at $z = 2.2$, 2.3 and 3.1, respectively.

\section{OBSERVATIONS \& Target fields}
\subsection{SPIRE OBSERVATIONS}

\begin{table*}
 \centering
 \caption{Summary of our  \textit{Herschel}/SPIRE observations.}
 \scalebox{0.88}[0.88]{
   \begin{tabular}{cccccccccccc}
   \hline
     Target & $z^a$ & R.A.$^b$ &Dec$^c$ & Area$^d$ & $t_{\rm{int}}$$^e$ & \multicolumn{3}{|c|}{$\sigma_{\rm{conf}}$$^f$} & \multicolumn{3}{|c|}{$\sigma_{\rm{inst}}$$^g$} \\ \smallskip
     & & & & & & $S_{250}$ & $S_{350}$ & $S_{500}$ & $S_{250}$ & $S_{350}$ & $S_{500}$  \\ 
      &  & (J2000) & (J2000) & (arcmin$^2$) & (hours) & (mJy) & (mJy) & (mJy) & (mJy) & (mJy) & (mJy) \\
     \hline \hline
     2QZCluster & 2.230 $\pm$ 0.016 & 10h03m51s & $+$00d15m09s & 515 & 1.8 &6.7 &7.0 &7.1 &2.0--3.9 &1.6--2.6 &2.0--3.3 \\ 
     HS1700 & 2.300 $\pm$ 0.015 & 17h01m15s & $+$64d14m03s & 497 & 1.5 &7.3 &7.4 &7.5  &2.0--3.7 &1.6--2.7 &2.0--3.2 \\ \smallskip
     SSA22 & 3.09 $\pm$ 0.03 & 22h17m34s & $+$00d17m01s & 1076 & 3.7 &7.7 &7.8 &8.0  &1.9--3.2 &1.6--2.4 &1.9--3.0 \\
     COSMOS & - & 10h00m37s & $+$02d11m26s & 3422 & 50.1 &7.1 &7.7 &7.8  &2.1--2.7 &1.7--2.0 &2.1--2.8 \\ \hline
     \end{tabular}}
     \\
\begin{flushleft}
{\bf Notes.} $(a)$: Redshift range of the member galaxies, based on HAEs for 2QZCluster \citep{2011MNRAS.416.2041M} and LBGs for HS1700 \citep{2005ApJ...626...44S} and SSA22 \citep[][]{1998ApJ...492..428S, 2000ApJ...532..170S}. $(b), (c)$: The coordinates of the field centre of \textit{Herschel}/SPIRE observations. $(d)$: The area where the integration time is greater than 30\%. For COSMOS field, we used centre of 3422 arcmin$^2$ area. We detected sources within this area. (see also Figure~3). $(e)$: The total integration time. For the survey design of COSMOS field, please see \citet{2012MNRAS.424.1614O}. $(f), (g)$: The confusion noise and instrumental noise.
\end{flushleft}
\end{table*}

Our \textit{Herschel}/SPIRE observations were performed as part of the second Open Time (OT2)  \textit{Herschel} programs (PI: Y. Matsuda). We summarize the observations in Table~1.  The observations were executed in Large Map mode with a scan rate of 30$''\ \rm{s}^{-1}$, repeated 14 times for each field ($N_{\rm{rep}}$=14). The dates of observations are 22 June 2012 (2QZCluster), 4 March 2012 (HS1700), and 10 May 2012 (SSA22). The coverage of the maps are $\sim$ $23'\times23'$ (2QZCluster), $\sim$ $22'\times22'$ (HS1700), and $\sim$ 33$'\times33'$ (SSA22) corresponding to $\sim40-60$ comoving Mpc at the protocluster redshifts, which are sufficient to search for concentration of DSFGs around the density peak of protocluster members. The integration times are 1.8, 1.5 and 3.7 hours for 2QZCluster, HS1700 and SSA22, respectively.  Maps were produced with the  \textit{Herschel} Interactive Processing Environment (HIPE, v11.0.0), following the standard data processing and map-making steps with destriping.  The full width at half maximum (FWHM) of the SPIRE beam is 18.1$''$, 24.9$''$ and 36.6$''$ at 250, 350 and 500~$\mu$m, respectively \citep{2010A&A...518L...4S}. The final maps have pixel sizes of 6$''$, 10$''$, and 14$''$ at 250, 350, and 500~$\mu$m. We measured the 1$\sigma$ confusion noise ($\sigma_{\rm{conf}}$) as map variance of flux density within effective area (Table~1), which are slightly higher than the values of blank fields \citep{2010A&A...518L...5N}. The instrumental noise measured in our three  protoclusters are about one-third of the confusion noise (Table~1).


\begin{figure*} 
 \centering
\includegraphics[scale=0.32]{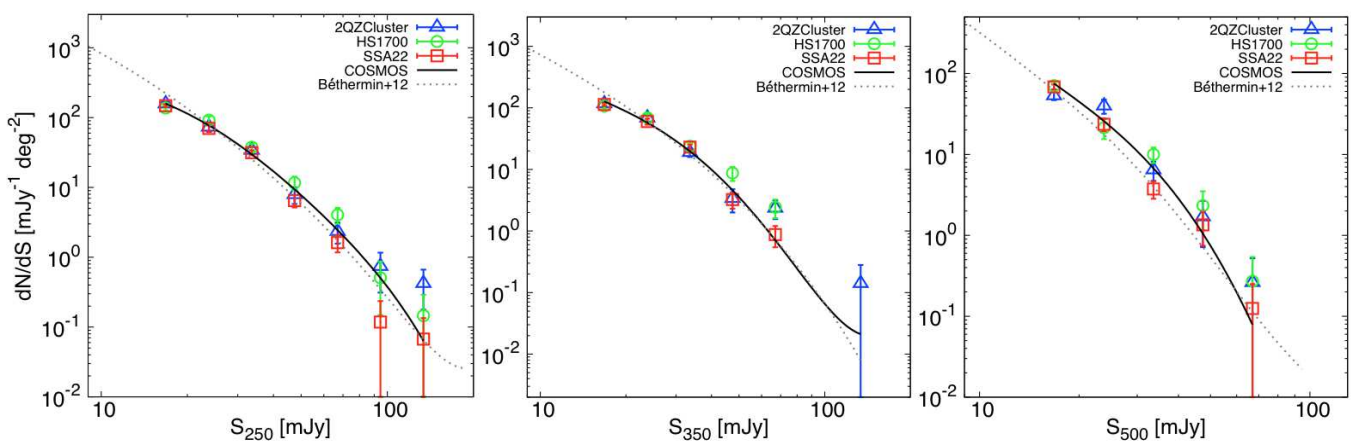}
 \caption{Number counts at 250, 350 and 500~$\mu$m in the three protoclusters and COSMOS field. We detected S/N $>$ 2 sources in the 250~$\mu$m maps and measured the 350 and 500~$\mu$m fluxes at the positions of the 250~$\mu$m sources. We fitted the data points of COSMOS filed with a b{\'e}zier curve. The number counts of the SPIRE sources in the protoclusters averaged over their full fields are roughly consistent with those of COSMOS field. The number counts of the SPIRE sources in COSMOS field are also consistent with those of HerMES blank fields \citep{2012A&A...542A..58B}, suggesting that COSMOS field can be used as a control field.}
\end{figure*}

\subsection{PROTOCLUSTER TARGETS}
\noindent{\bf 2QZCluster}; This protocluster was originally identified as a concentration of five Quasi Stellar Objects (QSOs) in a $\sim$ 1 degree region at $z=2.23$ from the 2dF Quasar Redshift survey \citep{2001MNRAS.322L..29C, 2004MNRAS.349.1397C}. Four out of the five QSOs are even more strongly clustered in a $30 \times 30$~comoving Mpc patch. An H$\alpha$ narrow-band imaging revealed a filamentary large-scale structure of an over density of 22 HAEs connecting the QSOs \citep{2011MNRAS.416.2041M}.  \textit{Chandra}/ACIS-I 100~ks observations of this structure also showed evidence that the Active Galactic Nucleus (AGNs) fraction is a factor of $\sim 3.5$ higher than blank fields \citep{2013ApJ...765...87L}. 

\smallskip
\noindent{\bf HS1700}; This protocluster was originally discovered as a $\sim7\times$ density contrast redshift spike of UV/optical-selected star forming galaxies (BX/BM) within a $\sim$ 25 comoving Mpc region at $z=2.30$ \citep{2005ApJ...626...44S}.  A Ly$\alpha$ narrowband imaging survey revealed a filamentary large-scale structure of six giant Ly$\alpha$ Blobs (LABs) \citep{2011ApJ...740L..31E}.   \textit{Chandra}/ACIS-I 200~ks observations of this structure also showed tentative evidence of an enhancement of AGN fraction compared to the field environment \citep{2010MNRAS.407..846D}.  


\smallskip
\noindent{\bf SSA22}; This protocluster was originally discovered as a $\sim4-6\times$ density contrast in redshift distribution of Lyman Break Galaxies (LBGs) and Ly$\alpha$ emitters (LAEs) within a $\sim 20$~comoving Mpc region at $z=3.09$ \citep[][]{1998ApJ...492..428S, 2000ApJ...532..170S}.  A wide-field Ly$\alpha$ narrowband imaging survey with Subaru Telescope revealed a filamentary large-scale structure of 283 LAEs, extended to at least $\sim60$~comoving Mpc \citep[][]{2004AJ....128.2073H}. This sample of LAEs includes 35 LABs with sizes of 30--150 kpc scale \citep{2000ApJ...532..170S, 2004AJ....128..569M}.  \textit{Chandra}/ACIS-I 400~ks observations of this structure showed that the AGN fraction of protocluster members is $\sim 3 \times$ higher than that in the field environment \citep[][]{2009ApJ...691..687L, 2009MNRAS.400..299L}. 

\smallskip
\noindent{\bf Blank field (COSMOS)}; We have chosen well-studied extragalactic field Cosmic Evolution Survey (COSMOS) to use as a blank field, which is observed as a part of  \textit{Herschel} Multi-tiered Extra-galactic survey, HerMES \citep{2012MNRAS.424.1614O}. The SPIRE map in COSMOS field is larger and deeper than our observations, so we have reprocessed them by limiting to the same depth using $N_{\rm{rep}}=14$. Subsequently, we apply the same map making, and source detection procedure in order to make sure that we match the depth of our observations. The confusion and instrumental noise values are given in Table~1.

\section{ANALYSIS}


\begin{table*}
\centering
  \caption{Summary of  \textit{Herschel}/SPIRE sources in the three protoclusters.}
  \scalebox{0.85}[0.85]{
   \begin{tabular}{cccccccccc}\hline
     Field & Area$^a$ & N$^b$ & N$^c$ & N$^d$ & N$^e$ & $\Sigma$$^f$& $\Sigma$$^g$ & $\delta$$^h$ & $\sigma$$^i$ \\
     & (arcmin$^{-2}$) & (prior) & (catalogued) & (selected) & & (arcmin$^{-2}$) & (arcmin$^{-2}$) & & \\ \hline \hline
     2QZCluster & 515 & 643 & 399 & 12 &  6 & 0.023 $\pm$ 0.007  & 0.132 $\pm$ 0.054 & 3.0 $\pm$ 1.7 & 3.9 $\pm$ 2.1 \\   
     HS1700 & 497 & 579 & 383 & 26 & 8 & 0.052 $\pm$ 0.010 &  0.186 $\pm$  0.066 & 3.7 $\pm$ 1.7 & 5.0 $\pm$ 2.2 \\ 
     SSA22 & 1076 & 1253 & 772 &  55 & 5 & 0.051 $\pm$ 0.007 & - & -  & - \smallskip \\ 
          COSMOS (matched to 2QZCluster) & 3422 & 4923 & 2777 & 111 & 7 & 0.032 $\pm$ 0.003 &-& - & - \\
        COSMOS (matched to HS1700) & 3422 & 4923 & 2777 & 140 & 7 & 0.041 $\pm$ 0.004 &-& - & - \\
      COSMOS (matched to SSA22) & 3422 & 4923 & 2777 & 262 & 10 & 0.077 $\pm$ 0.005 &-& - & - \\ \hline
   \end{tabular}}
   \begin{flushleft}
{\bf Notes.} $(a)$: The area where the integration time is greater than 30\%. We detected sources within this area. $(b)$: The numbers of S/N (250~$\mu$m) $>$ 2 prior sources. $(c)$:  The numbers of catalogued sources which at least one SPIRE band flux is above 12~mJy. $(d)$: The numbers of colour-selected bright SPIRE sources. For SSA22, we excluded the sources in high background fluxes (see Figure~3). $(e)$: The numbers of colour-selected bright SPIRE sources within the overdensities. $(f)$: The surface density of colour-selected bright SPIRE sources. The errors assume Poisson statistics. $(g)$: The surface density of colour-selected bright SPIRE sources in the overdensities, assume an  area of the overdensities ($r=3.8'$ for 2QZCluster, $r=3.7'$ for HS1700). The errors assume Poisson statistics. $(h)-(i)$: The $\delta$ and $\sigma$ are calculated from $\delta_{\rm{pc}}=(n_{\rm{pc}}-n_{\rm{ave}})/n_{\rm{ave}}$ and $\sigma_{\rm{pc}}=(n_{\rm{pc}}-n_{\rm{ave}})/\sigma_{\rm{ave}}$. The errors assume Poisson statistics. 
   \end{flushleft}
\end{table*}


\begin{figure*}
\center
\includegraphics[scale=0.33]{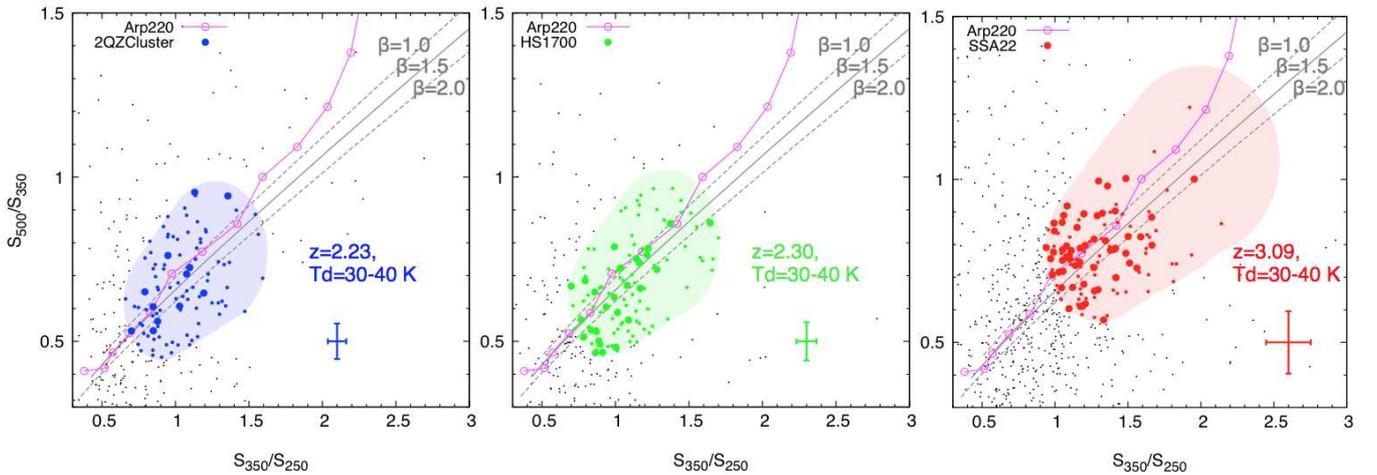}
\caption{$S_{500}/S_{350}$ vs $S_{350}/S_{250}$ colour-colour diagram of the  250~$\mu$m sources. From left to right, we show the plots for 2QZCluster, HS1700, and SSA22. We selected candidates of DSFGs possibly associated with the protoclusters whose colours are consistent with a single gray body SEDs including a photometric error of $\pm\ 20$\% (shaded regions). We assume the protocluster redshifts, dust temperatures of $T_d=30-40$ K and dust emissivity $\beta=1.5$. The grey solid and dashed lines show tracks of single grey body SEDs for different $\beta$. We plot sources with fluxes above 12mJy in least one SPIRE bands as small grey points. The coloured larger/smaller symbols show sources with $L_{\rm{FIR}}$ larger/smaller than $5.0\times 10^{12}\ L_\odot$. The error bars show the average errors of colour-selected bright SPIRE sources. We have selected 2\%, 5\%, and 4\% colour-selected bright SPIRE sources to search for overdensities of DSFGs in 2QZCluster, HS1700, and SSA22. We plot the expected colours for Arp 220 as a function of redshifts \citep{2007ApJ...663...81P} for comparison. We plot open circles on track of Arp220 every $z=0.5$ from $z=0$ to $z=5.5$.}
\end{figure*}


\begin{figure*}
\center
\includegraphics[scale=0.22]{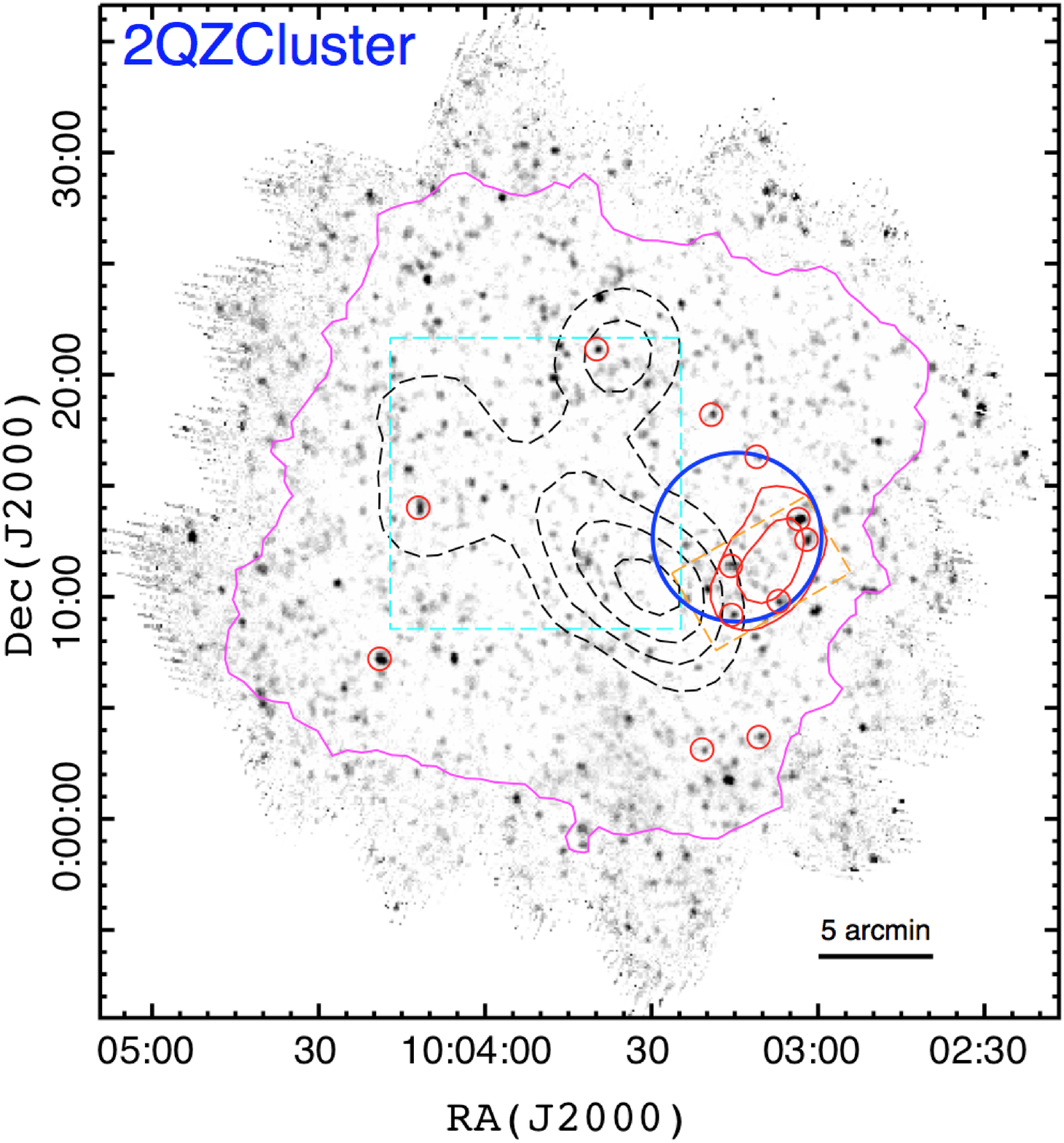}
\includegraphics[scale=0.228]{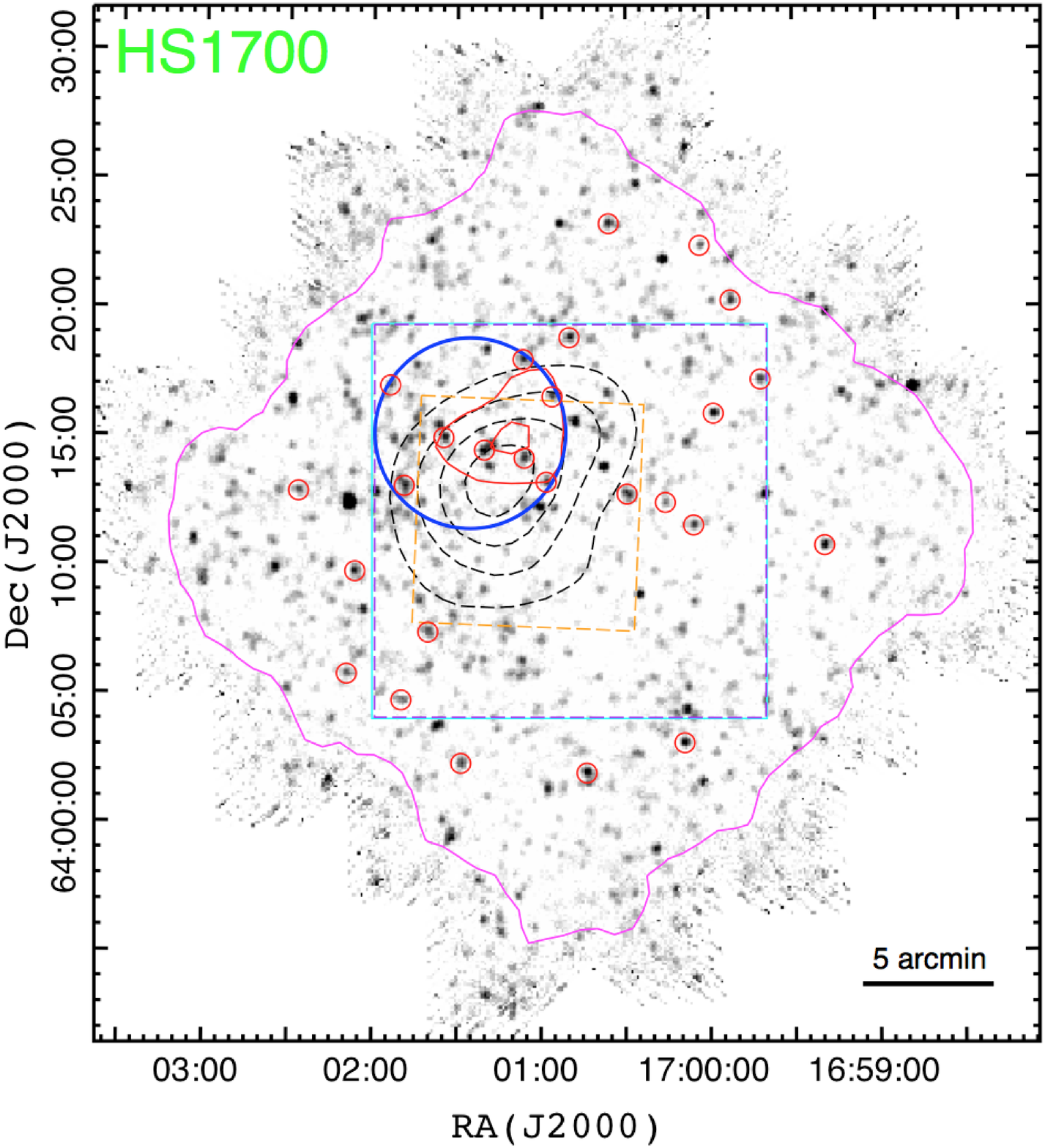}
\includegraphics[scale=0.3]{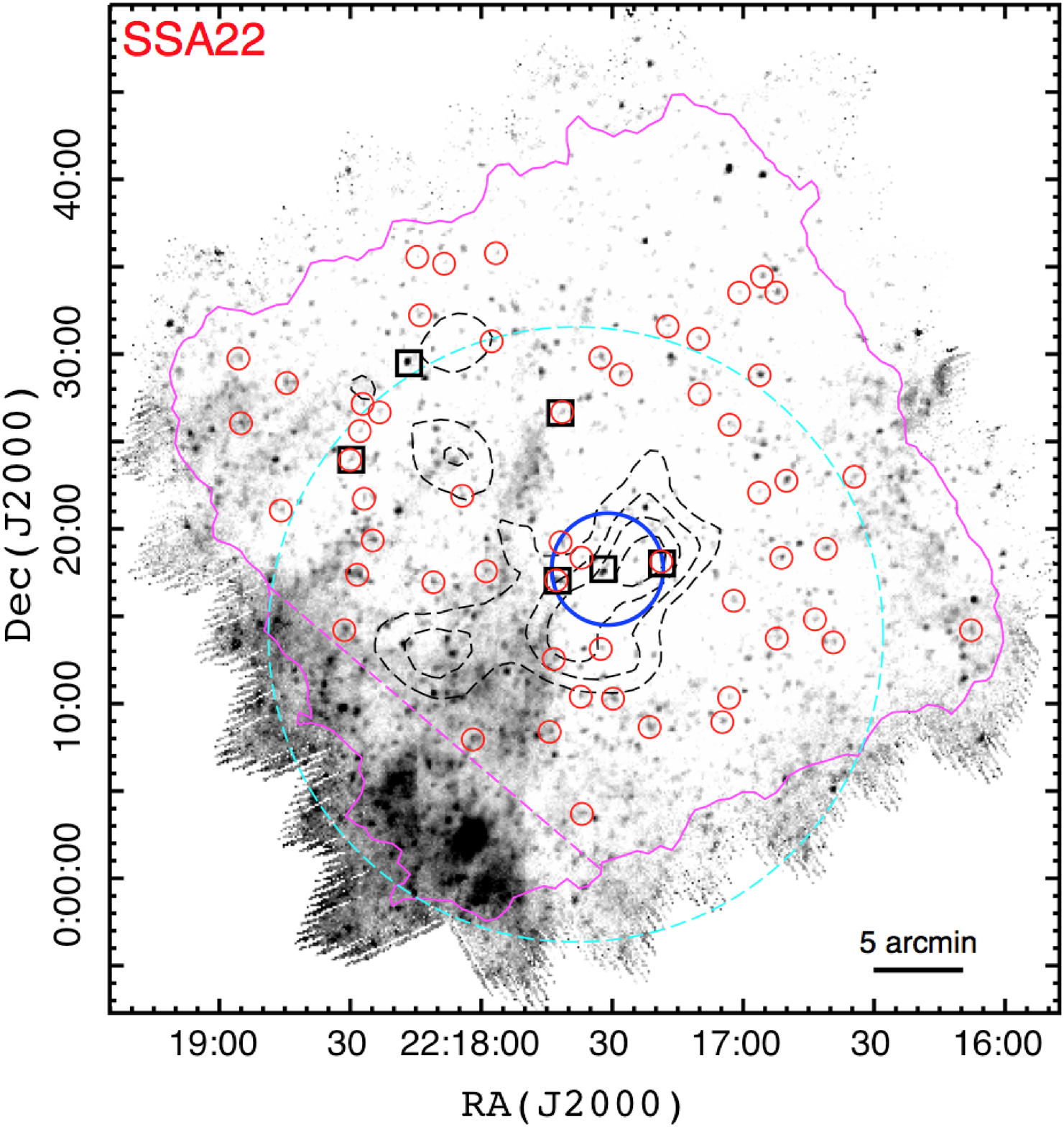}
\caption{Sky distributions of colour-selected bright 250~$\mu$m sources (red circles) over-plotted on the SPIRE 250~$\mu$m maps. The areas outlined with magenta contour corresponds to the 30\% depth coverage. For SSA22, we do not use below the dashed magenta line due to the high background fluxes. The dashed black contours show the density of 19  HAEs and 3 QSOs \citep{2011MNRAS.416.2041M} for the 2QZCluster, 45 LBGs for HS1700 \citep{2012ApJ...750...67R, 2014ApJ...795..165S}, and 742 LAEs for SSA22 \citep[][]{2012AJ....143...79Y} respectively. The steps show 1--4$\sigma$, 1--4$\sigma$, and 3--6$\sigma$ for 2QZCluster, HS1700, and SSA22 (smoothed with a Gaussian kernel with a FWHM of 6 comoving Mpc). The red contours show 3--4$\sigma$, 4--5$\sigma$ for colour-selected bright SPIRE sources in 2QZCluster and HS1700 with a same Gaussian kernel ($1\sigma$ is standard deviation of surface density of colour-selected bright SPIRE sources measured in COSMOS field). The large blue circles show the overdensity of colour-selected bright 250~$\mu$m sources for 2QZCluster and HS1700, and colour-selected bright 500~$\mu$m sources for SSA22 using filtering a radius of 6 comoving Mpc. We find a 4$\sigma$ and 5$\sigma$ overdensity in the 2QZCluster and HS1700 fields. We do not find any significant overdensities of 250~$\mu$m sources in SSA22, but we found six colour-selected bright 500~$\mu$m sources (black squares), and three sources are concentrated $3'$ ($\sim$1.4 Mpc) east to the LAEs overdensity. The dashed cyan and orange boxes in 2QZCluster show the UKIRT/WFCAM (for HAEs) and Subaru/MOIRCS (for HAEs) coverage. The solid cyan, dashed purple, and orange boxes in HS1700 show the Keck/LRIS (for LAEs and LBGs, respectively) and Palomar/WIRC (for HAEs)  coverage. The dashed cyan large circle in SSA22 shows ASTE/AzTEC coverage.}
\end{figure*}

\subsection{DETECTION \& NUMBER COUNTS}
The source detection was conducted on the 250\,$\mu$m maps, because of the better spatial resolution compared with longer wavelength bands. We cation that 500~$\mu$m detection causes a source blending (even 250~$\mu$m can deblend these) and large spatial uncertainty. We used the {\sc sussextractor} \citep{2012MNRAS.419..377S} for the source detection and photometry.  We detected S/N $>$ 2 sources in the maps within the region where the integration time is greater than 30\% of the deepest parts (515 arcmin$^2$ for 2QZCluster, 497 arcmin$^2$ for HS1700, 1076 arcmin$^2$ for SSA22, and 3422 arcmin$^2$ for COSMOS field). We measured 350~$\mu$m and 500~$\mu$m fluxes at the positions of sources detected in the 250~$\mu$m maps, then listed these 250~$\mu$m sources only if the flux density is above 12~mJy in at least one of the SPIRE bands (See also Table~8--10, and the full tables are avaivable in online). This flux density limit corresponds to $\sim 4\sigma$ of the instrumental noise and $\sim 2\sigma$ of the confusion noise in all three bands. For SSA22, we cut out the region shown in Figure~3 due to the high background fluxes from the Galactic cirrus. 

We compared these SPIRE number counts with COSMOS field. We show the raw (i.e., not corrected for the completeness) number counts in Figure~1 (See also Table~5-- 7). The raw number counts are roughly consistent with COSMOS field data at $>20$~mJy bin at 250 and 350~$\mu$m. A moderate excess of number counts at $S_{\rm{350}}>50$~mJy and $S_{\rm{500}}>40$ mJy were found (by a factor of 2--3) in HS1700 and 2QZCluster, although they are within the error bars based on Poisson noise. We also compared number counts in COSMOS field with wider  HerMES survey data \citep{2012A&A...542A..58B}. The number counts in COSMOS field agree with that of HerMES survey data,  suggesting that COSMOS field is suitable as a control field.

\subsection{SPIRE COLOUR SELECTION}
In order to search for DSFGs possibly associated with the protoclusters, we applied a colour constraint to the SPIRE detected sources. The $S_{350}/S_{250}\ \rm{and}\ S_{500}/S_{350}$ colours were used to select the sources with single grey body SEDs at the protocluster's redshift ranges, with dust temperatures of $T_d= 30-40$~K, and a dust emissivity $\beta=1.5$. \citet{2012ApJ...761..140C} measured the dust temperatures of SPIRE-selected DSFGs and showed that they are in the range of $T_d =  20-60$~K. The dust emissivity index is determined by \citet{1983QJRAS..24..267H}, and typically $\beta = 1.0 - 2.0$ for DSFGs. Different values of $\beta$ do not significantly affect our conclusion, yielding almost the same temperature and SPIRE colours. We included flux errors of $\pm\ 20\%$ in estimating the uncertainties in the SPIRE colours. Figure~2 shows $S_{500}/S_{350}\ \rm{vs.}\ S_{350}/S_{250}$ colour-colour diagram of the SPIRE sources in the three protoclusters.

We derived $L_{\rm{FIR}}$ ($8-1000~\mu$m) by fitting single grey body SEDs with $T_d$ a free parameter. From the sources selected based on the SPIRE colours, we further applied a FIR luminosity cut of $L_{\rm{FIR}} \geq 5.0\times 10^{12}\ L_{\odot}$ for conservative searching for DSFGs. Thus the lowest flux densities in our samples are (36.0, 30.0, 19.4)~mJy for 2QZCluster, (28.8, 26.3, 15.2)~mJy for HS1700, and (15.6, 19.2, 12.9)~mJy for SSA22 in the SPIRE bands (250, 350, 500)~$\mu$m,  respectively.
We finally obtained a sample of colour-selected bright SPIRE sources with colours consistent with the protocluster redshifts by rejecting $\sim95$\% of the 250~$\mu$m sources (thus we select just 12/643 (2\%), 26/579 (5\%), and 55/1253 (4\%) colour-selected bright SPIRE sources in 2QZCluster, HS1700 and SSA22). For COSMOS field, we applied the same colour selection as the three protoclusters to obtain a control sample of field galaxies. We have selected 111/4923 (2\%), 140/4923 (3\%), and 262/4923 (5\%) colour-selected bright SPIRE sources in COSMOS field  for 2QZCluster, HS1700, and SSA22 (see also Table~2).

\section{RESULTS}
\subsection{SEARCH FOR OVERDENSITIES}

\begin{figure}
\center
\includegraphics[scale=0.223]{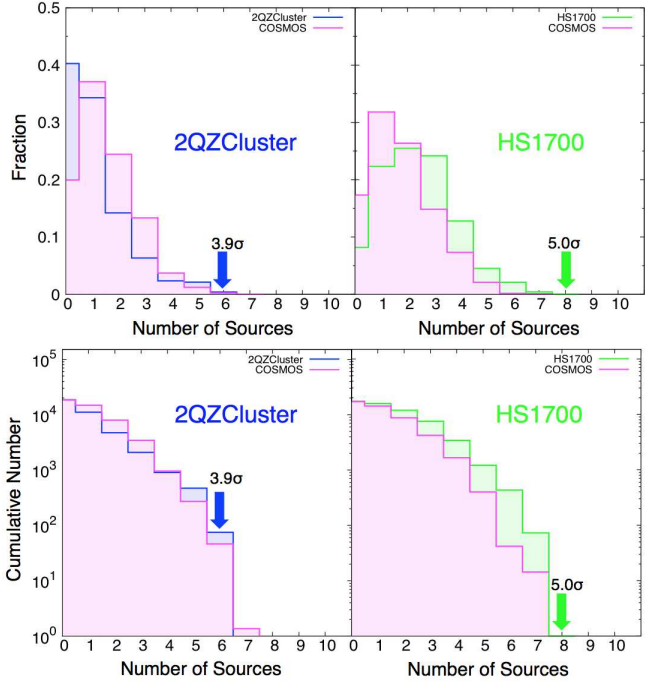}
\caption{(\textit{Upper}). The density distributions of the colour-selected bright SPIRE sources in 2QZCluster (left) and HS1700 (right) within an aperture radius of 6 comoving Mpc. 2QZCluster tends to have lower number of colour-selected bright SPIRE sources compared with COSMOS field. This could be explained if there are void-like structures within the survey area although there are also overdense regions. In contrast, HS1700 tends to have higher  number of sources. (\textit{Lower}). Cumulative number of the apertures with a radius of 6 comoving Mpc. We normalized the cumulative number of COSMOS field to the protoclusters.}
\end{figure}

We searched for overdensities in an aperture with a radius of 6~comoving Mpc (3.8$'$, 3.7$'$, and 3.2$'$ radius for $z=2.2, 2.3$, and $3.1$, respectively). This scale corresponds to physical scale of $\sim 1.5-2$~Mpc radius for each protocluster redshift, matching the size of the overdensity of DSFGs around radio galaxies \citep[e.g.,][]{2014MNRAS.437.1882R, 2014A&A...570A..55D}. We put down a grid of these apertures every 10$''$ for the protoclusters and COSMOS field, and counted the number of colour-selected bright SPIRE sources ($L_{\rm{FIR}} \geq 5.0\times 10^{12}\ L_{\odot}$) within each aperture. For SSA22, we excluded a high background region (see in Figure~3).

We found 3.9$\sigma$ and 5.0$\sigma$ overdensities in the 2QZCluster and HS1700 fields respectively, but did not find any significant ($>3\sigma$) overdensities in the SSA22 field. We calculated the significance of overdensity, $\sigma_{\rm{pc}}=(n_{\rm{pc}}-n_{\rm{ave}})/\sigma_{\rm{ave}}$, where $n_{\rm{pc}}$ is the number of colour-selected bright SPIRE sources in overdense region (6~comoving Mpc search radius) of the protocluster fields, $n_{\rm{ave}}$ is the average number within a 6 comoving Mpc search radius for COSMOS field, and $\sigma_{\rm{ave}}$ is the standard deviation of $n_{\rm{ave}}$. We also searched for overdensities of colour-selected bright SPIRE sources in case for 10\% and 30\% flux error boundaries in Figure~2. The result does not significantly change for 20\% and 30\% boundaries. We found 30\% boundary shows same position of overdensities of colour-selected bright SPIRE sources compared with that of 20\% boundary in 2QZCluster and HS1700. We found 10\% boundary is not suitable because the number of colour-selected bright SPIRE sources are too small to search for overdensities. In Figure~3, we show the sky distribution of the colour-selected bright SPIRE sources. We also show the density of 19 HAEs and 3 QSOs \citep{2011MNRAS.416.2041M} for 2QZCluster, 45 LBGs for HS1700 \citep{2012ApJ...750...67R, 2014ApJ...795..165S}, and 742 LAEs for SSA22 \citep[][]{2012AJ....143...79Y} respectively.

We searched for overdensities of the colour-selected bright SPIRE sources in the COSMOS field in an identical manner as for the protocluster fields. We plot the density distribution of the colour-selected bright SPIRE sources for the  2QZCluster, HS1700 and COSMOS fields in Figure~4. The histogram shows the distribution of the number of colour-selected bright SPIRE sources within a 6 comoving Mpc aperture (normalized by the  number of searched apertures). We found that 2QZCluster hasa low number of sources. The fraction of the apertures which have no sources is about two times higher compared with COSMOS field. This suggests that there is a void like distribution while there are also  overdensities in 2QZCluster. In contrast, HS1700 tends to have larger number of colour-selected bright SPIRE sources in the search apertures.

We also plot the cumulative number of the cricles in Figure~4. We normalized the cumulative number of COSMOS field to the protoclusters. We found that the normalized cumulative number with $>3.9\sigma$ overdensities in 2QZCluster field is about two times higher than that in the COSMOS field. There are no overdensities in the COSMOS field which contain eight colour-selected bright SPIRE sources ($5.0\sigma$) as HS1700. This means that such overdensities are preferentially located in the protoclusters.


\subsection{2QZCLUSTER}
In the 2QZCluster field, an overdensity of colour-selected bright SPIRE sources were found 4.5$'$ ($\sim$2.2 Mpc) west to the HAEs overdensity. 

We searched for the counterparts of SPIRE sources with HAEs and QSOs in 2QZCluster, which are summarized by \citet{2011MNRAS.416.2041M}, although the SPIRE overdense region did not have H$\alpha$ image with UKIRT/WFCAM (see Figure~3). In total, 19 HAEs and 3 QSOs are within the 174 arcmin$^2$ overlap region. We define the following quality criteria based on \citet{1986MNRAS.218...31D} for assessing the robustness of identified candidate counterparts. We classify sources with $p\leq0.05$ as secure counterparts, and those with $0.05 < p \leq 0.10$ as tentative counterparts. We calculated the $p$-value defined by $p=1-\rm{exp}(-\pi n\theta^2)$ where $n$ is the source density, $\theta$ is the angular offset. We searched for counterparts within 11$''$ radius from the centre of the SPIRE sources. This search radius corresponds to $\sim$40\% beam response in 250~$\mu$m band. 

For the QSOs, 1 optically luminous QSO, which is known as 2QZC-C1-HAE2 \citep{2011MNRAS.416.2041M}, coincides with one of our colour-selected bright SPIRE source 2QZCluster-SPIRE10 (2.7$''$ offset). The $p$-value is lower than 0.01, suggesting a secure counterpart, the probability of chance association of counterparts is lower than 1\%. 2QZ-C1-HAE3 also coincides with our SPIRE source 2QZCluster-SPIRE124 (5.6$''$ offset) as a secure counterpart. Such optically luminous FIR bright QSOs are thought to be in a transient phase between DSFGs and QSOs \citep{2012MNRAS.426.3201S}. No other HAEs matched with colour-selected bright SPIRE sources. We then repeated this matching using all 250~$\mu$m sources, and found five secure HAE counterparts  (see Table~11).

We recently conducted Subaru/MOIRCS follow-up NB imaging observations in May 5, 2015 to search for new HAEs as well as counterparts of the SPIRE sources in the SPIRE overdense region (4$'\times7'$; dashed orange box in Figure 3). The total exposure time of $K_{\rm{s}}$ and NB data is each 1.5 and 2.4 ksec, respectively. The 5$\sigma$ detection limit of NB $<$ 19.8 (Vega) is almost the same as the previous HAE search depth (NB $<$ 19.9 in Vega) by \citet{2011MNRAS.416.2041M}.

We find no new HAE within the area. This may suggest that the area is actually a little under-dense in terms of bright HAE, because \citet{2011MNRAS.416.2041M} claim an average field count of 0.09 per arcmin$^2$, which corresponds to two or three HAEs per MOIRCS FOV. Such void structures adjacent to protocluster has been reported previously \citep[e,g.,][]{2013MNRAS.428.1551K, 2015MNRAS.447.3069S}.

\subsection{HS1700}
In HS1700, the overdensity peak of colour-selected bright SPIRE sources coincides with that of the protocluster member galaxies (2.1$'$ offset, corresponding to $\sim$ 1.0 Mpc). 

We searched for the counterparts of SPIRE sources using UV-selected (BX/BM) star-forming galaxies (LBGs), HAEs, LAEs, and DRGs in the SPIRE overdense region. There are 3010 LBGs within 256 arcmin$^2$, 123 LAEs are within 219 arcmin$^2$, 83 HAEs and 75 DRGs are within 72 arcmin$^2$. The LBG catalogues are from the KBSS \citep[Keck Baryonic Structure Survey;][]{2012ApJ...750...67R, 2014ApJ...795..165S}. HS1700 is one of the fifteen QSO fields intensively studied in KBSS. The H$\alpha$ and Ly$\alpha$ NB catalogues are from Milan Bogosavljevi\'c’s Ph.D thesis \citep{2010PhDT.......343B}. Individual properties of LBGs, LAEs and observations are described in \citet{2005ApJ...626..698S}, \citet{2011ApJ...740L..31E}, \citet{2013ApJ...774..130K}, and \citet{2014ApJ...795...33E}. In addition, for the H$\alpha$ narrow band imaging, we used the Br$\gamma$ filter (Palomar WIRC, center wavelength 2.17 $\mu$m, FWHM=297 \AA ). The narrow band selection criteria are NB $<20.5$ (Vega) and NB $-K_s \leq  -0.75$. The DRGs have selected with $J-K>2.3$ to a limit of $K_s=21$ (Vega). We define the quality criteria, and classify sources with $p$-value described above. We deduced the number density of 45 LBGs which have spectroscopically confirmed redshift in the protocluster (i.e., in the overdensity of $z=2.285-2.315$) separately from remaining 2965 LBGs outside of overdensity. We searched for counterparts within 11$''$ radius from the centre of the SPIRE sources.

We found three colour-selected bright SPIRE sources that have secure counterparts (see Table~12). But one of the LAEs BNB1, the spectrum of this object is extremely odd and that makes difficult to identify the redshift. We identified it as a “Lo-BAL” QSO and it is more likely to have $z\sim$ 2.00, so we do not treat as it is a protocluster member galaxy.

We also found one LAE and LBG are secure, and one DRG and LAE are tentative counterparts of colour-selected faint (i.e., $L_{\rm{FIR}} < 5.0\times 10^{12}\ L_\odot$) SPIRE source. For the SPIRE sources which are not selected with our colour selection, but have at least one SPIRE band above 12 mJy, we found two secure LAEs, three secure HAEs, and one tentative HAE. Thus, we conclude that seven SPIRE sources in the overdensity have secure counterparts of protocluster galaxies (see Table~12). 

In order to assess the success rate of the SPIRE colour selection, we compared the matching result for colour-selected bright/faint SPIRE sources and all 250~$\mu$m sources lying within overdensity. We assumed that all HAEs and LAEs are associated with the protocluster except for BNB1, although we should note that the HAEs without spectroscopic follow-up are much less likely than spectroscopic confirmed HAEs to be at the cluster redshift. For all secure counterparts, the fractions are 2/7(29\%), 1/5(20\%), and 4/28(14\%), respectively. This suggests that our SPIRE colour selection can select possible protocluster members with three times higher probability compared to not colour-selected sources.

In the HS1700 field, there are several foreground galaxy groups in the field. A  $z=0.453$ group is very close to the position of HS1700-SPIRE24. Indeed,  \citet{2007ApJ...668...23P} found that BX913 is lensed by the group. We calculated a magnification factor by using {\sc glafic} \citep{2010PASJ...62.1017O}. We used a halo mass of $M/h=1 \times10^{14}\ M_{\odot}$ \citep{2014A&A...564A.129I} and a concentration parameter of c=6  \citep{2013ApJ...766...32B}. We found that the second nearest (1.5$'$ away from group center) SPIRE source is affected only $\sim$5\% magnification, so we conclude that only the HS1700-SPIRE24 could be affected by lensing.

\subsection{SSA22}
The maximum number of colour-selected bright SPIRE sources found in on 6 comoving Mpc aperture was five, corresponding to a $1.6\sigma$ overdensity. This suggests that SSA22 does not have any significant overdensities of colour-selected bright 250~$\mu$m sources. But, this could be a problem for SSA22 at $z=3.1$ because DSFGs would start to drop out at 250~$\mu$m. Here we searched for an overdensity of colour-selected bright 500~$\mu$m detected sources with identical colour selection and luminosity cut in \S3.2.  We found six colour-selected bright 500~$\mu$m sources in the SSA22 field,  and three sources are concentrated $3'$ ($\sim$1.4 Mpc) east to the LAEs overdensity (Figure~3). For 350~$\mu$m sources, we did not find any significant overdensities as same as 250~$\mu$m sources. 

We investigated AzTEC 1.1 mm counterparts in \citet{2014MNRAS.440.3462U,2015ApJ...815L...8U} for the five colour-selected bright 500~$\mu$m sources within $\sim$800 arcmin$^2$ overlapped region (Figure~3). We searched for the counterparts within a radius of 14$''$ (half of AzTEC 1.1 mm FWHM). We found that four colour-selected bright 500~$\mu$m sources have been matched (SSA22-AzTEC1, 2, 5, and 34). Thus $\sim80$\% colour-selected bright 500~$\mu$m sources are matched with AzTEC 1.1 mm sources.

\section{DISCUSSION \& SUMMARY}
We searched for DSFGs associated with three protoclusters at $z=2-3$ (2QZCluster, HS1700, SSA22) using \textit{Herschel}/SPIRE. In the 2QZCluster and HS1700 field, we found $4\sigma$ and $5\sigma$ overdensities of the colour-selected bright SPIRE sources on a scale of 6~comoving Mpc. In the SSA22, we did not find any significant overdensities of 250~$\mu$m sources, but we found that three colour-selected bright 500~$\mu$m sources are concentrated on a scale of 6~comoving Mpc. The results suggest possible activity associated with enhanced dusty star-formation in protoclusters at $z\sim2-3$. 


\begin{figure*}
\center
\includegraphics[scale=0.28]{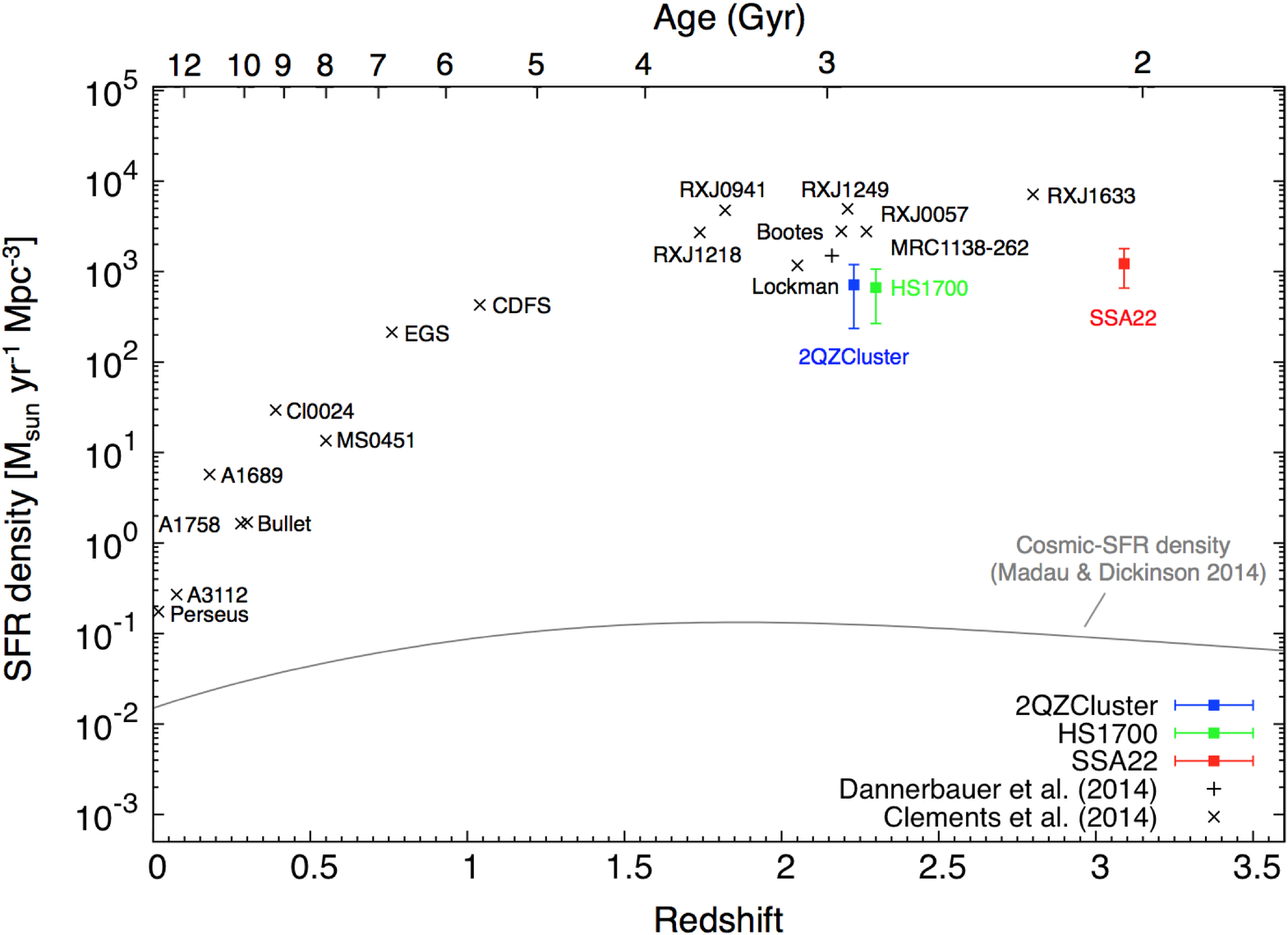}
\caption{SFR density of clusters, protoclusters and global cosmic SFR densities. The SFR densities of our protoclusters are 10$^3$--10$^4$ times higher than the global SFR density \citep{2014ARA&A..52..415M}. We show the compilation from \citet{2014A&A...570A..55D} and data from \citet{2014MNRAS.439.1193C}, which original literature are following; \textit{IRAS} measurements of Perseus from \citet{2000A&A...363..933M}, BLAST measurements of A3112 from \citet{2011MNRAS.412.1187B}, \textit{ISO} measurements of A1689 from \citet{2000A&A...361..827F}, and \textit{Spitzer} measurements of A1758 from \citet{2009MNRAS.396.1297H}, Bullet cluster from \citet{2010ApJ...725.1536C}, Cl0024+16 and MS0451-03 from \citet{2006ApJ...649..661G}. RXJ0057, RXJ0941, RXJ1218, RXJ1249 and RXJ1633 are based on JCMT/SCUBA from \citet{2010MNRAS.405.2623S}.}
\end{figure*}

We derive the star-formation rate (SFR) density of 2QZCluster, HS1700 and SSA22 to compare with the average value of the Universe. The SFR of DSFGs is often converted from far-infrared luminosity, although the conversion from $L_{\rm{FIR}}$ to SFR is not straightforward and relies on the dust composition and initial mass function (IMF). Most works on DSFGs assume the conversion given by $\rm{SFR}\ (M_{\odot}\ \rm{yr}^{-1}) = 4.5 \times\ 10^{-44}\ \it{L}_{\rm{FIR}}\ \rm{(erg\ s^{-1})}$ or $1.7 \times\ 10^{-10}\ \it{L}_{\rm{FIR}}\ (\it{L}_{\odot})$ \citep{1998ARA&A..36..189K}, where $\it{L}_{\rm{FIR}}$ is the integrated luminosity of $8-1000~\mu$m. This conversion takes the radiative transfer models of \citet{1995ApJS...96....9L} and a Salpeter IMF \citep{1955ApJ...121..161S}.

We used a simple estimation to calculate the integrated SFR densities by assuming that all colour-selected bright SPIRE sources in the overdensities are associated with protoclusters. In addition, we also included sources with S/N $>$ 2 in all three SPIRE bands within a radius of 1~Mpc (physical) following \citet{2014MNRAS.439.1193C} analysis of \textit{Planck} clumps with  \textit{Herschel}/SPIRE. Because this radius of 1~Mpc circles are smaller than our search radius (6~comoving Mpc, $\sim1.5-2$~Mpc in physical), we adjusted the position of the smaller aperture to contain as many colour-selected bright SPIRE sources as possible. This enables us to select regions similar to \textit{Planck} clumps in \citet{2014MNRAS.439.1193C}. The $1\sigma$ instrumental noise of Clements's work ranges from $2.5-2.8$ mJy at 250~$\mu$m, $2.1-2.3$ mJy at 350~$\mu$m and $3.0-3.3$ mJy at 500~$\mu$m \citep{2014MNRAS.439.1193C}, which is similar to our survey. For all SPIRE sources within a 1~Mpc radius, we fitted single grey body SEDs with fixed $T_d=35$~K, $\beta=1.5$ and the protocluster's redshifts, and derived $L_{\rm{FIR}}$. We excluded one colour-selected bright SPIRE source HS1700-SPIRE30 from this discussion because it is very close to $z=0.08$ SDSS galaxy, and its \textit{Spitzer}/MIPS 24 $\mu$m flux density is very high ($\sim200\ \mu$Jy). We found that Mips 24 $\mu$m flux densities of other colour-selected bright SPIRE sources in HS1700 are consistent with that of typical high-$z$ star-forming galaxies. 2QZCluster does not have any \textit{Spitzer} images.

Figure~5 shows our results and compare to previous studies in the literature. We calculated SFR with \citet{2003ApJ...586..794B}'s $L_{\rm{FIR}}$ correction to follow \citet{2014MNRAS.439.1193C} analysis. The derived SFR densities are 10$^3$--10$^4$ times higher than the global SFR density seen in \citet{2014ARA&A..52..415M}. This comparison indicates enhanced star-formation activities in protoclusters at $z\sim2-3$. The enhancement of dusty star-formation activity within our protoclusters is consistent with that seen for \textit{Planck} clumps \citep{2014MNRAS.439.1193C}. \citet{2014A&A...570A..55D} presents  results based on APEX/LABOCA 870~$\mu$m observations around  MRC1138$-$262 at $z=2.16$, showing that at least six DSFGs are likely part of the protocluster. The SFR density $\sim 1500$~$\rm{M_{\odot}}$~yr$^{-1}$~Mpc$^{-3}$ is similar to \citet{2014MNRAS.439.1193C} and our protoclusters study. We also plot the error bars in Figure~5. The high end shows the case when including all sources in 1 Mpc radius and the low end shows the case for subtracting field average values. We deduced field values by calculating average number of sources and average luminosity in 1 Mpc radius apertures in the COSMOS field. 

In Table~3, we summarized the estimated SFR and assumed volume plotted in Figure~5. We also caluculated SFR density in an aperture using a radius of 6 comoving Mpc in the  same manner. These radii are 1.9, 1.8 and 1.5 Mpc in physical units for 2QZCluster, HS1700 and SSA22, respectively, and the results do not change compared to the case of 1 Mpc.


\begin{table}
\center
  \caption{Estimated SFR and assumed volume for Figure 5.}
 \scalebox{0.78}[0.78]{
   \begin{tabular}{ccccccc}\hline
      Field & $z$ & SFR$^a$ & $r^b$ & $V^c$ & SFR$_{\rm{min}}$$^d$\\ 
      &  & ($\rm{M_{\odot}}$ yr$^{-1}$) & (Mpc) & (Mpc$^3$) &  ($\rm{M_{\odot}}$ yr$^{-1}$) \\ \hline \hline
                     \multicolumn{6}{|c|}{All sources} \\ \\
     2QZCluster & $2.230\pm0.016$ & 5000 & 1.0 & 4.2 & 150\\ 
     HS1700 & $2.300\pm0.015$ & 4500 & 1.0 & 4.2 & 280\\ \smallskip \smallskip 
     SSA22 & $3.09\pm0.03$ & 7500 & 1.0 & 4.2 & 310\\
               \multicolumn{6}{|c|}{Field corrected values} \\ \\
                    2QZCluster & $2.230\pm0.016$ & 1000 & 1.0 & 4.2 & - \\ 
     HS1700 & $2.300\pm0.015$ & 2000 & 1.0 & 4.2 & - \\
     SSA22 & $3.09\pm0.03$ & 4700 & 1.0 & 4.2 & - \\ \hline
   \end{tabular}}
   \begin{flushleft}
   {\bf Notes.} $(a)$: Integrated SFR derived from SPIRE sources which are included in column 4 radius. $(b)-(c)$: Used radius and volume to calculate SFR density. $(d)$: Minimum SFR of SPIRE source which is included in integrated SFR. 
   \end{flushleft}
\end{table}

What causes the enhanced dusty star-forming activity in our protoclusters? One possible answer to this is galaxies mergers. For instance, \citet{2015ApJ...808L..33C} investigated the  morphology of the galaxies in a protocluster to search for interaction/merger state by using \textit{HST} H-band imaging data.  Although the sample size is limited, they found that the fraction of irregular and interacting galaxies among the LBGs and DSFGs is 1.5 times higher in the protocluster than in the field.  \citet{2015ApJ...809..173W} suggested that dusty star formation at the center of a $z=1.7$ cluster is being driven by galaxy-galaxy interaction, involving a far-infrared luminous Bright Cluster Galaxy (BCG). Because there are strong correlation between the major-merger and the far-infrared luminosity \citep{2010ApJ...721...98K, 2012ApJ...757...23K, 2010ApJ...724..233E, 2013ApJ...778..129H}, higher major mergers rate in protoclusters could induce dusty star-formation. N-body simulation also predicts that progenitors of cluster and group haloes at $z>2$ have $3-5 \times$ higher major merger rates than isolated halos \citep{2001ApJ...546..223G}. The facts that the colour-selected bright SPIRE sources are significantly concentrated in HS1700 and 2QZCluster would support such simultaneously major merging phenomenon. However, further observations are needed to investigate the processes of dusty star-forming activity in these protoclusters.

\smallskip
\section*{ACKNOWLEDGMENTS}
We thank the anonymous referee for helpful comments which significantly improved the clarity of this paper. 
We thank Scott Chapman, James Colbert, Emanuele Daddi, Koichiro Nakanishi and Kazuhiro Shimasaku for useful discussions and supports.
Rhythm Shimakawa and Mariko Kubo made enormous contribution to analyses.
We acknowledge Masaru Kajisawa for our use of his MOIRCS fringe-removal software \citep{2015ApJ...801..134K} during the data reduction.

\textit{Herschel} is an ESA space observatory with science instruments
provided by European-led Principal Investigator consortia and with
important participation from NASA. SPIRE has been developed by a consortium of institutes led by
Cardiff University (UK). Subaru Telescope, which is operated by the National Astronomical Observatory of Japan.
UKIRT is funded by the STFC (UK). The W.M. Keck Observatory was made possible by the generous
financial support of the W.M. Keck Foundation. The authors wish
to recognize and acknowledge the very significant cultural role and
reverence that the summit of Mauna Kea has always had within
the indigenous Hawaiian community. We are most fortunate to
have the opportunity to conduct observations from this mountain.

This research was supported in part by a grant from the Hayakawa Satio Fund awarded by the Astronomical Society of Japan. YM acknowledges support from JSPS KAKENHI Grant Number 20647268. IRS acknowledges support from STFC (ST/L00075X/1), the ERC Advanced Grant DUSTYGAL (321334) and a Royal Society/Wolfson Merit Award. BH acknowledges support from JSPS KAKENHI Grant Number 15K17616. HU acknowledges support from Grant-in-Aid for JSPS Fellows, 26.11481.







\begin{table*}
\center
  \caption{Number of sources which have fluxes above 12 mJy in each bands.}
 \scalebox{1.0}[1.0]{
   \begin{tabular}{cccccc}\hline
      &  COSMOS & SSA22 & HS1700 & 2QZCluster \\ \hline \hline
     250~$\mu$m & 2552 & 699 & 356 & 365  \\ 
     350~$\mu$m & 1944 & 529 & 259 & 260 \\
     500~$\mu$m & 961 & 252 & 133 & 116 \\ \hline
   \end{tabular}
   }
\end{table*}

\begin{table*}
\center
  \caption[Number counts at 250~$\mu$m.]{Number counts at 250~$\mu$m. The errors take into account the statistical uncertainties, including the Poisson noise.}
 \scalebox{1.0}[1.0]{
   \begin{tabular}{ccccccc}\hline
     Central flux &  Flux bin &  \multicolumn{4}{c}{Number counts ($dN/dS$) [mJy$^{-1}$ deg$^{-2}$] }  \\
     (mJy)& (mJy) & SSA22 & HS1700 & 2QZCluster & COSMOS  \\ \hline \hline
     16.8& 12.0 -- 21.6 & 148.1 $\pm$ 7.2 & 137.2 $\pm$ 10.2 & 157.1 $\pm$ 10.7 & 158.1 $\pm$ 4.2 \\ 
     23.8& 21.6 -- 26.0& 70.0 $\pm$ 7.3 & 92.1 $\pm$ 12.3 & 73.0 $\pm$ 10.8 & 87.5 $\pm$ 4.6 \\ 
     33.6 & 26.0 -- 41.2 & 31.5 $\pm$ 2.6 & 37.6 $\pm$ 4.2 & 34.0 $\pm$ 4.0 & 36.4 $\pm$ 1.6 \\   
     47.4 & 41.2 -- 53.6 &6.5 $\pm$ 1.3 & 11.7 $\pm$ 2.6 & 7.9 $\pm$ 2.1 & 10.4 $\pm$ 0.9 \\
     67.0 & 53.6 -- 80.4& 1.6 $\pm$ 0.5 & 4.0 $\pm$ 1.0 & 2.3 $\pm$ 0.8 & 2.7 $\pm$ 0.3 \\
     94.6 & 80.4 -- 108.8& 0.12 $\pm$ 0.12 & 0.51 $\pm$ 0.36 & 0.74 $\pm$ 0.43 & 0.74 $\pm$ 0.17 \\
     133.7 & 108.8 -- 158.6& 0.07 $\pm$ 0.07  & 0.15 $\pm$ 0.15 & 0.42 $\pm$ 0.24 & 0.06 $\pm$ 0.04  \\
     188.8 & 158.6 -- 219.0 &- & - & - & -  \\ \hline
   \end{tabular}
   }
\end{table*}

\begin{table*}
\center
  \caption[Number counts at 350~$\mu$m.]{Number counts at 350~$\mu$m. The errors take into account the statistical uncertainties, including the Poisson noise.}
 \scalebox{1.0}[1.0]{
   \begin{tabular}{ccccccc}\hline
     Central flux & Flux bin &  \multicolumn{4}{c}{Number counts ($dN/dS$) [mJy$^{-1}$ deg$^{-2}$] }  \\
     (mJy)& (mJy) & SSA22 & HS1700 & 2QZCluster & COSMOS \\ \hline \hline
     16.8& 12.0 -- 21.6 & 113.6 $\pm$ 6.3 & 107.0 $\pm$ 9.0 & 116.4 $\pm$ 9.2 & 127.4 $\pm$ 3.7  \\ 
     23.8& 21.6 -- 26.0 & 60.1 $\pm$ 6.8 & 69.1 $\pm$ 10.7 & 68.3 $\pm$ 10.4 & 62.2 $\pm$ 3.9  \\ 
     33.6 & 26.0 -- 41.2 & 23.1 $\pm$ 2.3 & 23.8 $\pm$ 3.4 & 18.8 $\pm$ 2.9 & 28.5 $\pm$ 1.4  \\   
     47.4 & 41.2 -- 53.6 & 3.2 $\pm$ 0.9 & 8.8 $\pm$ 2.3 & 3.4 $\pm$ 1.4 & 5.9 $\pm$ 0.7  \\
     67.0 & 53.6 -- 80.4 & 0.9 $\pm$ 0.3 & 2.4 $\pm$ 0.8 & 2.3 $\pm$ 0.8 & 1.5 $\pm$ 0.2  \\
     94.6 & 80.4 -- 108.8 & - & - & - & 0.04 $\pm$ 0.04 \\
     133.7 & 108.8 -- 158.6 & - & - & 0.14 $\pm$ 0.14 & - \\
     188.8 & 158.6 -- 219.0 & - & - & - & -  \\ \hline
   \end{tabular}
   }
\end{table*}

\begin{table*}
\center
  \caption[Number counts at 500~$\mu$m.]{Number counts at 500~$\mu$m. The errors take into account the statistical uncertainties, including the Poisson noise.}
 \scalebox{1.0}[1.0]{
   \begin{tabular}{ccccccc}\hline
     Central flux & Flux bin &  \multicolumn{4}{c}{Number counts ($dN/dS$) [mJy$^{-1}$ deg$^{-2}$] }  \\
     (mJy)& (mJy) & SSA22 & HS1700 & 2QZCluster & COSMOS  \\ \hline \hline
     16.8& 12.0 -- 21.6 & 68.3 $\pm$ 4.9 & 70.8 $\pm$ 7.3 & 53.1 $\pm$ 6.2 & 75.2 $\pm$ 2.9 \\ 
     23.8& 21.6 -- 26.0 & 23.6 $\pm$ 4.2 & 21.4 $\pm$ 5.9 & 39.7 $\pm$ 7.9 & 27.3 $\pm$ 2.6  \\ 
     33.6 & 26.0 -- 41.2 & 3.7 $\pm$ 0.9 & 10.0 $\pm$ 2.2 & 6.4 $\pm$ 1.7 & 9.7 $\pm$ 0.8  \\   
     47.4 & 41.2 -- 53.6 & 1.3 $\pm$ 0.6 & 2.3 $\pm$ 1.2 & 1.7 $\pm$ 1.0 & 1.6 $\pm$ 0.4 \\
     67.0 & 53.6 -- 80.4 & 0.1 $\pm$ 0.1 & 0.27 $\pm$ 0.27 & 0.26 $\pm$ 0.26 & 0.08 $\pm$ 0.06  \\
     94.6 & 80.4 -- 108.8 & - & - & - & -  \\
     133.7 & 108.8 -- 158.6 & - & - & - & - \\
     188.8 & 158.6 -- 219.0 & - & - & - & -  \\ \hline
   \end{tabular}
   }
\end{table*}

\begin{table*}
\center
  \caption{SPIRE sources catalogue of 2QZCluster. The full table is available online.}
 \scalebox{1.0}[1.0]{
   \begin{tabular}{ccccccc}\hline
ID&R.A.&Dec.& $S_{\rm{250}}$ & $S_{\rm{350}}$ & $S_{\rm{500}}$ \\
&(J2000) & (J2000) & (mJy) & (mJy) & (mJy) \\ \hline \hline
2QZCluster-SPIRE1 & 150.81846968368112 & 0.029424092103377656 & 127.0 $\pm$ 2.1 & 54.2 $\pm$ 1.9 & 24.6 $\pm$ 2.2 \\
2QZCluster-SPIRE2 & 150.76480950482065 & 0.22501254698200338 & 139.1 $\pm$ 2.4 & 117.9 $\pm$ 1.8 & 71.4 $\pm$ 2.3 \\
2QZCluster-SPIRE3 & 150.91517959664077 & 0.3909833462929028 & 111.4 $\pm$ 2.4 & 55.2 $\pm$ 1.8 & 21.2 $\pm$ 2.4 \\
2QZCluster-SPIRE4 & 151.02463506680982 & 0.1206541823404277 & 97.4 $\pm$ 2.2 & 54.2 $\pm$ 1.8 & 26.5 $\pm$ 2.4 \\
2QZCluster-SPIRE5 & 151.04455762206453 & 0.4047339937359634 & 93.6 $\pm$ 2.3 & 61.3 $\pm$ 1.8 & 24.2 $\pm$ 2.2 \\
2QZCluster-SPIRE6 & 151.07930718745527 & 0.12006911913976963 & 84.1 $\pm$ 2.3 & 58.9 $\pm$ 1.8 & 31.3 $\pm$ 2.3 \\
2QZCluster-SPIRE7 & 151.11225122198036 & 0.17674219593981788 & 66.1 $\pm$ 2.2 & 36.2 $\pm$ 2.1 & 13.2 $\pm$ 2.1 \\
2QZCluster-SPIRE8 & 150.7066091154151 & 0.2830585502030584 & 67.4 $\pm$ 2.3 & 54.5 $\pm$ 1.8 & 18.4 $\pm$ 2.2 \\
2QZCluster-SPIRE9 & 150.94968775985714 & 0.33116978814976 & 64.4 $\pm$ 2.2 & 22.2 $\pm$ 1.9 & 0.0 $\pm$ 2.1 \\
2QZCluster-SPIRE10 & 150.91631977926835 & 0.3524558479517708 & 60.1 $\pm$ 2.1 & 64.7 $\pm$ 1.8 & 45.6 $\pm$ 2.0 \\ \hline
   \end{tabular}
   }
\end{table*}

\begin{table*}
\center
  \caption{SPIRE sources catalogue of HS1700. The full table is available online.}
 \scalebox{1.0}[1.0]{
   \begin{tabular}{ccccccc}\hline
ID &R.A. &Dec. & $S_{\rm{250}}$ & $S_{\rm{350}}$ & $S_{\rm{500}}$ \\
&(J2000) & (J2000) & (mJy) & (mJy) & (mJy) \\ \hline \hline
HS1700-SPIRE1 & 255.53646683056928 & 64.2060083253161 & 422.6 $\pm$ 3.1 & 181.4 $\pm$ 1.9 & 67.3 $\pm$ 2.4 \\
HS1700-SPIRE2 & 255.07094202680116 & 64.36337246187729 & 112.4 $\pm$ 2.2 & 49.0 $\pm$ 1.9 & 19.5 $\pm$ 2.2 \\
HS1700-SPIRE3 & 255.10511462429938 & 64.14686204692669 & 86.2 $\pm$ 2.4 & 51.5 $\pm$ 2.0 & 21.4 $\pm$ 2.1 \\
HS1700-SPIRE4 & 255.15806474143926 & 64.22916135482777 & 85.8 $\pm$ 2.5 & 60.5 $\pm$ 1.8 & 27.9 $\pm$ 2.2 \\
HS1700-SPIRE5 & 254.99437853070503 & 64.26410380634614 & 70.8 $\pm$ 2.1 & 76.4 $\pm$ 1.7 & 51.2 $\pm$ 2.2 \\
HS1700-SPIRE6 & 254.91863041158092 & 64.21174308418698 & 76.7 $\pm$ 2.4 & 44.4 $\pm$ 1.7 & 17.6 $\pm$ 2.3 \\
HS1700-SPIRE7 & 255.22359843297363 & 64.38646521955137 & 65.7 $\pm$ 2.2 & 21.3 $\pm$ 1.9 & 13.0 $\pm$ 2.3 \\
HS1700-SPIRE8 & 254.82938021291324 & 64.17873107989598 & 68.7 $\pm$ 2.3 & 59.4 $\pm$ 1.7 & 27.6 $\pm$ 2.2 \\
HS1700-SPIRE9 & 255.20168755944053 & 64.25859055126669 & 66.3 $\pm$ 2.2 & 35.3 $\pm$ 2.1 & 16.2 $\pm$ 2.2 \\
HS1700-SPIRE10 & 255.1823543841818 & 64.03153372522755 & 64.7 $\pm$ 2.2 & 64.8 $\pm$ 1.9 & 31.2 $\pm$ 2.1 \\ \hline
   \end{tabular}
   }
\end{table*}

\begin{table*}
\center
  \caption{SPIRE sources catalogue of SSA22. The full table is available online.}
 \scalebox{1.0}[1.0]{
   \begin{tabular}{ccccccc}\hline
ID &R.A. &Dec & $S_{\rm{250}}$ & $S_{\rm{350}}$ & $S_{\rm{500}}$ \\
&(J2000) & (J2000) & (mJy) & (mJy) & (mJy) \\ \hline \hline
SSA22-SPIRE1 & 334.264444128561 & 0.6764415586203842 & 135.3±2.2 & 71.3±1.9 & 30.4±2.3 \\
SSA22-SPIRE2 & 334.57160684240256 & 0.49219161422544583 & 93.3±2.2 & 76.7±1.7 & 48.7±2.3 \\
SSA22-SPIRE3 & 334.344178441303 & 0.3528652232847903 & 79.2±2.2 & 34.4±1.9 & 10.4±2.2 \\
SSA22-SPIRE4 & 334.34457572927244 & 0.607307593915126 & 79.7±2.4 & 56.9±1.7 & 21.9±2.0 \\
SSA22-SPIRE5 & 334.69745115684617 & 0.3983836901706686 & 72.9±2.2 & 29.8±1.9 & 12.8±2.1 \\
SSA22-SPIRE6 & 334.392819716858 & 0.3042332652494624 & 68.9±2.1 & 33.2±1.8 & 0.0±2.1 \\
SSA22-SPIRE7 & 334.23836274598824 & 0.40194158606064945 & 61.8±2.1 & 45.8±1.8 & 8.8±2.1 \\
SSA22-SPIRE8 & 334.55462830631444 & 0.47701863826781765 & 62.9±2.1 & 33.3±1.8 & 18.3±2.0 \\
SSA22-SPIRE9 & 334.3848999033597 & 0.2912395866246295 & 57.5±2.1 & 49.6±1.7 & 45.7±2.2 \\
SSA22-SPIRE10 & 334.58002216598203 & 0.35006798763473185 & 64.3±2.3 & 51.3±2.0 & 30.6±2.3 \\ \hline
   \end{tabular}
   }
\end{table*}


\begin{landscape}
\begin{table}
  \caption{Matching results for the SPIRE sources which have HAEs counterparts in 2QZCluster.}
  \scalebox{1.0}[1.0]{
   \begin{tabular}{cccccccccccc}\hline
     ID & R.A. & Dec & $S_{250}$ & $S_{350}$ & $S_{500}$ & Counterpart & R.A. & Dec & Sepa.$^{a}$ & $p$-value$^{b}$ & Member$^{c}$ \\
     (SPIRE) & (J2000) &  (J2000) & (mJy) & (mJy) & (mJy) &  & (J2000) & (J2000) & (arcsec) & & \\ \hline \hline \\
       \multicolumn{12}{|c|}{Colour-selected bright SPIRE sources} \\ \\ \smallskip \smallskip
2QZCluster-SPIRE10 & 150.916320 & 0.352456  & 60.1 $\pm$ 2.1 & 64.7 $\pm$ 1.8 & 45.6 $\pm$ 2.0 & QZC-C1-HAE02(QSO) & 150.915790 & 0.353000 & 2.73 & $<$0.01 & Secure \\ \hline \\ 
 \multicolumn{12}{|c|}{at least one SPIRE band above 12 mJy sources} \\ \\
2QZCluster-SPIRE124 & 150.965998 & 0.249463 & 21.9 $\pm$ 2.2 & 14.2 $\pm$ 1.8 & 9.7 $\pm$ 2.2 & QZC-C1-HAE03(QSO) & 150.964920 & 0.250583 & 5.60 & $<$0.01 & Secure \\
2QZCluster-SPIRE239 & 151.069470 & 0.310488 & 15.4 $\pm$ 2.2 & 4.1 $\pm$ 1.8 & 6.9 $\pm$ 2.3 & QZC-C1-HAE20 & 151.069750 & 0.309167 & 4.86 & $<$0.01 & Secure \\
2QZCluster-SPIRE251 & 150.885551 & 0.168325 & 15.5 $\pm$ 2.2 & 6.8 $\pm$ 2.0 & 3.7 $\pm$ 2.3 & QZC-C1-HAE09 & 150.887370 & 0.167917 & 6.71 & $<$0.01 & Secure \\
2QZCluster-SPIRE261 & 150.852526 & 0.155903 & 15.8 $\pm$ 2.3 & 13.8 $\pm$ 1.8 & 12.1 $\pm$ 2.2 & QZC-C1-HAE05 & 150.854380 & 0.155694 & 6.72 & $<$0.01 & Secure \\
2QZCluster-SPIRE353 & 150.916183 & 0.187292 & 14.4 $\pm$ 2.7 & 12.1 $\pm$ 1.8 & 3.0 $\pm$ 2.1 & QZC-C1-HAE16 & 150.915750 & 0.187722 & 2.20 & $<$0.01 & Secure \\
2QZCluster-SPIRE372 & 150.935267 & 0.238716 & 10.2 $\pm$ 2.2 & 16.2 $\pm$ 1.8 & 14.0 $\pm$ 2.1 & QZC-C1-HAE15 & 150.932670 & 0.238944 & 9.39 & 0.01 & Secure \\
\hline
   \end{tabular}}
   \begin{flushleft}
{\bf Notes.}  $(a)$: Separation between SPIRE source and counterparts. $(b)$:  $p$-value based on \citet{1986MNRAS.218...31D}. $(c)$: We classify sources with $p\leq0.05$ as secure, and those with $0.05 < p \leq 0.10$ as tentative counterparts. 
   \end{flushleft}
\end{table}
\end{landscape}


\begin{landscape}
\begin{table}
  \caption{Matching results for the SPIRE sources which have emitter counterparts within overdensity in HS1700.}
  \scalebox{1.0}[1.0]{
   \begin{tabular}{ccccccccccccc}\hline
     ID & R.A. & Dec & $S_{250}$ & $S_{350}$ & $S_{500}$& ID & R.A. & Dec & $z$$^{a}$  & Sepa.$^{b}$ & $p$-value$^{c}$ & Member$^{d}$ \\
     (SPIRE) & (J2000) &  (J2000) & (mJy) & (mJy) & (mJy) & (Count.) & (J2000) & (J2000) & & (arcsec) & & \\ \hline \hline \\
     \multicolumn{13}{|c|}{Colour-selected bright SPIRE sources} \\ \\
     HS1700-SPIRE21 & 255.3951 & 64.2484 & 52.5 $\pm$ 2.2 & 41.2 $\pm$ 1.7 &  21.1 $\pm$ 2.2 &  BNB1  & 255.3953 & 64.2479 & $\sim$2.00$^{e}$ & 1.63 & $<0.01$ & Secure\\ 
      & & & & & & BX980 & 255.3987 & 64.2490 & - & 6.13 & 0.32 & Not \\
     HS1700-SPIRE42 & 255.2428 & 64.2196 & 40.0 $\pm$ 2.3 & 46.4 $\pm$ 1.8 & 34.3 $\pm$  2.1 & DRG46 & 255.2419 &64.2195 & - & 1.37 & $<0.01$ & Secure\\
          & & & & & &  HaNB2 & 255.2417 & 64.2196 & - & 1.75 & $<0.01$ & Secure\\ 
     HS1700-SPIRE24 & 255.3349 & 64.2403 & 51.5 $\pm$ 2.3 & 63.0 $\pm$ 2.0 & 49.3 $\pm$ 2.2 & DRG53 & 255.3378 & 64.2395 & $\sim2.3^{f}$ & 5.36 & $ 0.03 $ & Secure\\
     & & & & & & HaNB10* & 255.3399 & 64.2384 & 2.289(abs) &10.43 & 0.10 & Tentative\\
     & & & & & & BX913* & 255.3399 & 64.2385 & 2.289(abs), 2.291(neb) &10.29 & 0.05 & Tentative\\ \smallskip \smallskip
     & & & & & & BX928 & 255.3318 & 64.2405 & 2.755(Ly$\alpha$) & 4.92 & 0.22 & Not\\  \hline \\
          \multicolumn{13}{|c|}{Colour-selected faint SPIRE sources} \\ \\
     HS1700-SPIRE321 & 255.2732 & 64.2046 & 13.7 $\pm$ 2.4 & 20.2 $\pm$ 1.8 & 15.9 $\pm$ 2.1 & BNB41* & 255.2688 & 64.2027 & 2.287(abs) & 9.64 & 0.05 & Secure \\ 
     & & & & & & MD109* & 255.2687 & 64.2026 & 2.293(abs,neb) & 10.09 & 0.02 & Secure \\ 
     & & & & & & DRG38$^+$ & 255.2669 & 64.2032 & 2.286(emi) & 10.98 & 0.10 & Tentative\\ \smallskip \smallskip 
     & & & & & & BNB16$^+$ & 255.2670 & 64.2033 & 2.290(emi) & 10.53 & 0.05 & Tentative \\ \hline \\
          \multicolumn{13}{|c|}{at least one SPIRE band above 12 mJy sources} \\ \\
     HS1700-SPIRE70 & 255.2827 & 64.2586 & 31.1 $\pm$ 2.2 & 21.4 $\pm$ 1.9 & 11.0 $\pm$ 2.0 & BNB155 & 255.2811 & 64.2574 & 2.290(Ly$\alpha$) & 5.13 & 0.01 & Secure\\
     & & & & & & HaNB27 & 255.2873 & 64.2578 & - & 7.81 & 0.06 & Tentative\\
HS1700-SPIRE78 & 255.4358 & 64.2607 & 30.3 $\pm$ 2.2 & 16.9 $\pm$ 1.7 & 6.0 $\pm$ 2.1 & BNB139 & 255.4331 & 64.2622 & - & 6.85 & 0.02 & Secure\\
HS1700-SPIRE142 & 255.2668 & 64.2469 & 22.7 $\pm$ 2.2 & 12.0 $\pm$ 1.8 & 1.8 $\pm$ 2.2 & HaNB83 & 255.2653 & 64.2487 & - & 6.67 & 0.04 & Secure\\
& & & & & & HaNB76 & 255.2661 & 64.2461 & - & 3.06 & 0.01 & Secure\\
HS1700-SPIRE179 & 255.3768 & 64.1988 & 19.8 $\pm$ 2.2 & 9.4 $\pm$ 1.8 & 0.5 $\pm$ 2.2 & HaNB45 & 255.3796 & 64.1996 & - &  5.41 &  0.03 & Secure\\
\hline
   \end{tabular}}
   \begin{flushleft}
   {\bf Notes.} $(a)$: Redshift information of counterparts. $(b)$: Separation between SPIRE source and counterparts. $(c)$: $p$-value based on \citet{1986MNRAS.218...31D}. $(d)$: We classify sources with $p\leq0.05$ as secure, and those with $0.05 < p \leq 0.10$ as tentative counterparts. $(e)$: BNB1 is identified as a “Lo-BAL” QSO and therefore it is difficult to measure redshift. $(f)$: Photometric redshift from \citet{2015MNRAS.449L..68C}.
   \end{flushleft}
\end{table}
\end{landscape}


\bsp	
\label{lastpage}
\end{document}